\documentclass[a4paper]{article}

\usepackage{INTERSPEECH2022}
\usepackage{multirow}
\usepackage{cite}
\usepackage{graphicx}
\usepackage{subfigure}

\title{A Hierarchical Speaker Representation Framework \\ for One-shot Singing Voice Conversion}
\name{Xu Li, Shansong Liu, Ying Shan}
\address{
  ARC Lab, Tencent PCG}
\email{\{nelsonxli, shansongliu, yingsshan\}@tencent.com}

\begin{document}

\maketitle

\begin{abstract}
Typically, singing voice conversion (SVC) depends on an embedding vector, extracted from either a speaker lookup table (LUT) or a speaker recognition network (SRN), to model speaker identity. However, singing contains more expressive speaker characteristics than conversational speech. It is suspected that a single embedding vector may only capture averaged and coarse-grained speaker characteristics, which is insufficient for the SVC task. To this end, this work proposes a novel hierarchical speaker representation framework for SVC, which can capture fine-grained speaker characteristics at different granularity. It consists of an up-sampling stream and three down-sampling streams. The up-sampling stream transforms the linguistic features into audio samples, while one down-sampling stream of the three operates in the reverse direction. It is expected that the temporal statistics of each down-sampling block can represent speaker characteristics at different granularity, which will be engaged in the up-sampling blocks to enhance the speaker modeling. Experiment results verify that the proposed method outperforms both the LUT and SRN based SVC systems. Moreover, the proposed system supports the one-shot SVC with only a few seconds of reference audio.

\end{abstract}

\section{Introduction}
Singing voice conversion (SVC) is a technology that transfers the voice of a singing signal to a voice of a target singer while maintaining the underlying content and melody.
It has been paid more and more attention due to its potential applications in human-computer interaction, entertainment, etc.

Singing signals contain many components mixed together, such as content, pitch, loudness, and speaker identity.
An SVC system should be able to disentangle these components, then reconstruct the converted singing with the target speaker information while keeping other components unchanged.
Prior work has explored adequately in disentangling singing components by separately encoding each component.
For the content component, Zhang et al. \cite{zhang2020durian} encodes the lyrics into hidden features to represent the content, while other approaches directly extract the content information from the source singing signal, e.g. extracting linguistic features by a pre-trained automatic speech recognition (ASR) model \cite{liu2021diffsvc,guo2020phonetic,polyak2020unsupervised}, applying adversarially trained content encoders \cite{deng2020pitchnet,nachmani2019unsupervised} and utilizing the vector quantized variational autoencoder (VQVAE) \cite{takahashi2021hierarchical,wang2021vqmivc}.
In terms of the pitch extraction, the log-F0 \cite{liu2021diffsvc,lu2020vaw} and sine-excitation \cite{chen2021singgan,guo2022improving} are two common methods to capture the pitch variation in singing signals.
Moreover, the A-weighting mechanism of the power spectrum \cite{liu2021fastsvc} and the root mean square energy \cite{zhang2020durian} are also adopted to measure the loudness for more precise SVC.

The speaker modeling in singing signals is non-trivial.
For a single singer conversion system, a.k.a. any-to-one conversion, it implicitly acquires the target speaker's characteristics from the training data \cite{liu2021diffsvc,chen2019singing}.
For a multi-singer system, speaker modeling is necessary for distinguishing different singers and facilitating singing voice conversion.
One of the commonly adopted methods in the multi-singer system training is the speaker lookup table (LUT) \cite{liu2021fastsvc,deng2020pitchnet,takahashi2021hierarchical}.
It assigns a fixed-dimensional vector to each singer in the training data.
The assigned vectors are learnable so as to adaptively capture the corresponding speaker characteristics during training.
The LUT-based approaches are typically leveraged in the any-to-many SVC scenario, however, it fails to perform SVC towards an unseen singer which is outside of the training data.
Other approaches \cite{zhang2020durian,guo2020phonetic,guo2022improving} leverage a speaker recognition network (SRN) to distill speaker information by extracting speaker embeddings from a target singer's reference audio.
This enables the one-shot SVC scenario where the target singer is not observed in the training data.

However, singing voice can be much more expressive than conversational speech.
The speaker characteristics in singing can greatly fluctuate even within a few seconds.
Probably the speaker information of a speech clip can, to some extend, be condensed into a compact speaker embedding.
But it is highly questionable that a single speaker embedding, extracted from either LUT or SRN, is adequate to capture the dynamically varying speaker characteristics of a singing utterance.
Thereby, this work proposes a novel SVC framework on exploring a hierarchical speaker representation to capture fine-grained speaker characteristics at different granularity.
Specifically, it adopts an up-sampling stream and three down-sampling streams.
The up-sampling stream contains several up-sampling blocks and transforms linguistic features into audio samples in an end-to-end manner.
Reversely, one down-sampling stream of the three alternates down-sampling blocks with instance normalization (IN) \cite{ulyanov2016instance} to transform audio samples back to linguistic features.
The IN module after each down-sampling block performs sample-wise normalization over the time dimension and generates temporal statistics for each down-sampling block, which can be utilized to represent speaker characteristics.
The temporal statistics of all the blocks form a hierarchical speaker representation at different granularity, which is fused into the up-sampling blocks to provide the speaker condition.

\begin{figure*}[th]
    \centering
    \includegraphics[width=\textwidth]{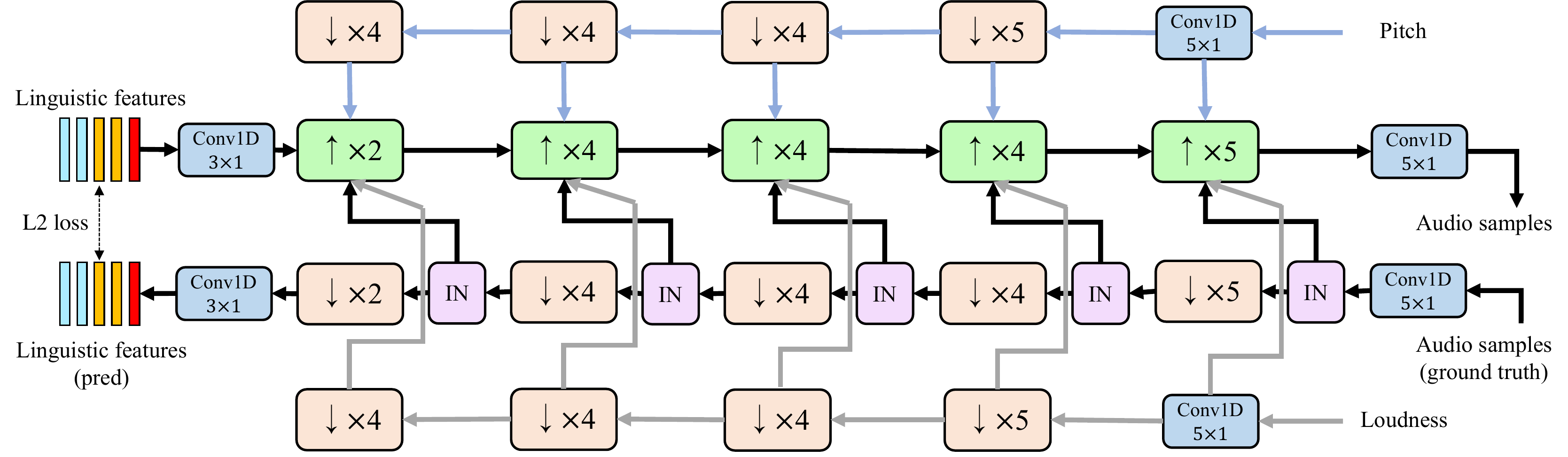}
    \caption{Schematic diagram of the proposed model architecture. In the figure, the ``$\uparrow$" and the ``$\downarrow$" represent the up-sampling and down-sampling blocks, respectively. The ``IN" denotes instance normalization.}
    \label{fig:proposed_framework}
\end{figure*}

A concurrent work \cite{li2021unet} on the text-to-speech task expresses a similar idea to ours, i.e. incorporating the IN module within a U-net structure to perform voice cloning.
The major differences between our present study and \cite{li2021unet} are:
1) \cite{li2021unet} absorbs both speaker and speaking style information, e.g. pitch, from the temporal statistics, while this work extracts only speaker information from the temporal statistics. The pitch and loudness information are explicitly fed into the system as done in \cite{liu2021fastsvc}. This can help disentangle speaker information from pitch and loudness; 
2) the U-net structure in \cite{li2021unet} transforms features between the content representations and the Mel-spectrograms. It utilizes a pre-trained MelGAN vocoder \cite{yang2021multi} for audio synthesis. 
While this work directly operates between content representations and audio samples in an end-to-end manner, which can alleviate the cumulative error in the multi-stage SVC \cite{guo2020phonetic}; 
3) Both this work and \cite{li2021unet} utilize a U-net-like architecture, but the block structures are different. \cite{li2021unet} designs the blocks as the AdaIN layers \cite{chou2019one} alternated with residual layers, while this work additionally adopts the feature-wise affine module \cite{chen2020wavegrad} to fuse other conditions into the up-sampling stream.

This work has the following contributions: 1) Proposing a novel SVC framework that captures a hierarchical speaker representation at different granularity; 2) Demonstrating the superior performance of the proposed system over the LUT and speaker embedding based systems; 3) Showing our system supporting the one-shot conversion with only a few seconds of reference audio \footnote{Audio samples: https://lixucuhk.github.io/Unet-SVC-Demo/}.

The rest of this paper is organized as follows: Section~\ref{sec:method} illustrates the proposed system architecture. Experimental setup and experiment results are demonstrated in Section~\ref{sec:expt-setup} and \ref{sec:expt-results}, respectively. We conclude this work in Section~\ref{sec:conclusions}.

\section{Proposed Method}
\label{sec:method}

\subsection{The proposed framework}
\label{subsec:proposed_framework}

The proposed SVC framework is illustrated in Fig.~\ref{fig:proposed_framework}. It is modified from the architecture of FastSVC \cite{liu2021fastsvc}, which is an end-to-end SVC framework with high efficiency. 
The difference is that this work presents a hierarchical speaker modeling at different granularity to provide speaker information, rather than utilizing a speaker LUT as that in \cite{liu2021fastsvc}.
As shown in Fig.~\ref{fig:proposed_framework}, the system consists of an up-sampling stream together with three down-sampling streams.
The up-sampling stream transforms linguistic features block by block into the audio samples.
The down-sampling stream with black arrows alternates down-sampling blocks with instance normalization (IN) \cite{ulyanov2016instance} to transform audio samples back to linguistic features.
The IN module has been shown promising in vision and speech tasks \cite{ulyanov2016instance,chou2019one}, it performs sample-wise normalization over width/height/time dimensions to extract sample-wise characteristics.
In this work, it performs sample-wise normalization over the time dimension to extract global temporal statistics, which can be utilized to represent speaker characteristics \cite{chou2019one}.
The temporal statistics of all the blocks form a hierarchical speaker representation at different granularity, which is fused into the up-sampling blocks to provide the fine-grained speaker information. This work utilizes the mean as the temporal statistics for speaker representation, which will be explained in Section~\ref{subsec:wav_gen}.
Moreover, the other two down-sampling streams operate on the pitch and loudness features, respectively.
The pitch and loudness features are explicitly extracted from the source singing voice, then up-sampled to match the rate of audio samples.
After that, they are down-sampled accordingly to match the rate of each up-sampling block, and fused into the up-sampling blocks.
Section~\ref{subsec:feat_extraction} will detail the feature extraction process, while the structure of the building blocks will be demonstrated in Section~\ref{subsec:wav_gen}.
Finally, the training loss will be discussed in Section~\ref{subsec:training-loss}.

\subsection{Feature extraction}
\label{subsec:feat_extraction}

\textbf{Linguistic feature extraction:} A hybrid CTC-attention ASR system \cite{gulati2020conformer} is adopted as the content extractor, which achieves state-of-the-art ASR performance.
It consists of a Conformer \cite{gulati2020conformer} based encoder, a CTC decoder and an attention decoder, and adopts 80-dimensional log Mel-spectrograms as the input features.
We follow the large version of the training configuration presented in Table 1 of \cite{gulati2020conformer} to train the system on the LibriSpeech corpus.
Both decoders are discarded after training, and 512-dimensional vectors are extracted from the last output layer of the encoder as the linguistic features.

\textbf{Pitch extraction:} The Log-F0 \cite{liu2021diffsvc,lu2020vaw} and the sine-excitation \cite{chen2021singgan,guo2022improving} are two common methods for pitch representation.
Previous studies \cite{guo2022improving,polyak2020unsupervised} have shown that harmonic sine-excitation signals can improve the smoothness and continuity of the generated singing voice, which are thus adopted in this work.
We extend the single sinusoid sine-excitation adopted in \cite{liu2021fastsvc} to multiple harmonics \cite{chen2021singgan} for more robust pitch representation.

\begin{figure}[th]
    \centering
    \includegraphics[width=0.45\textwidth]{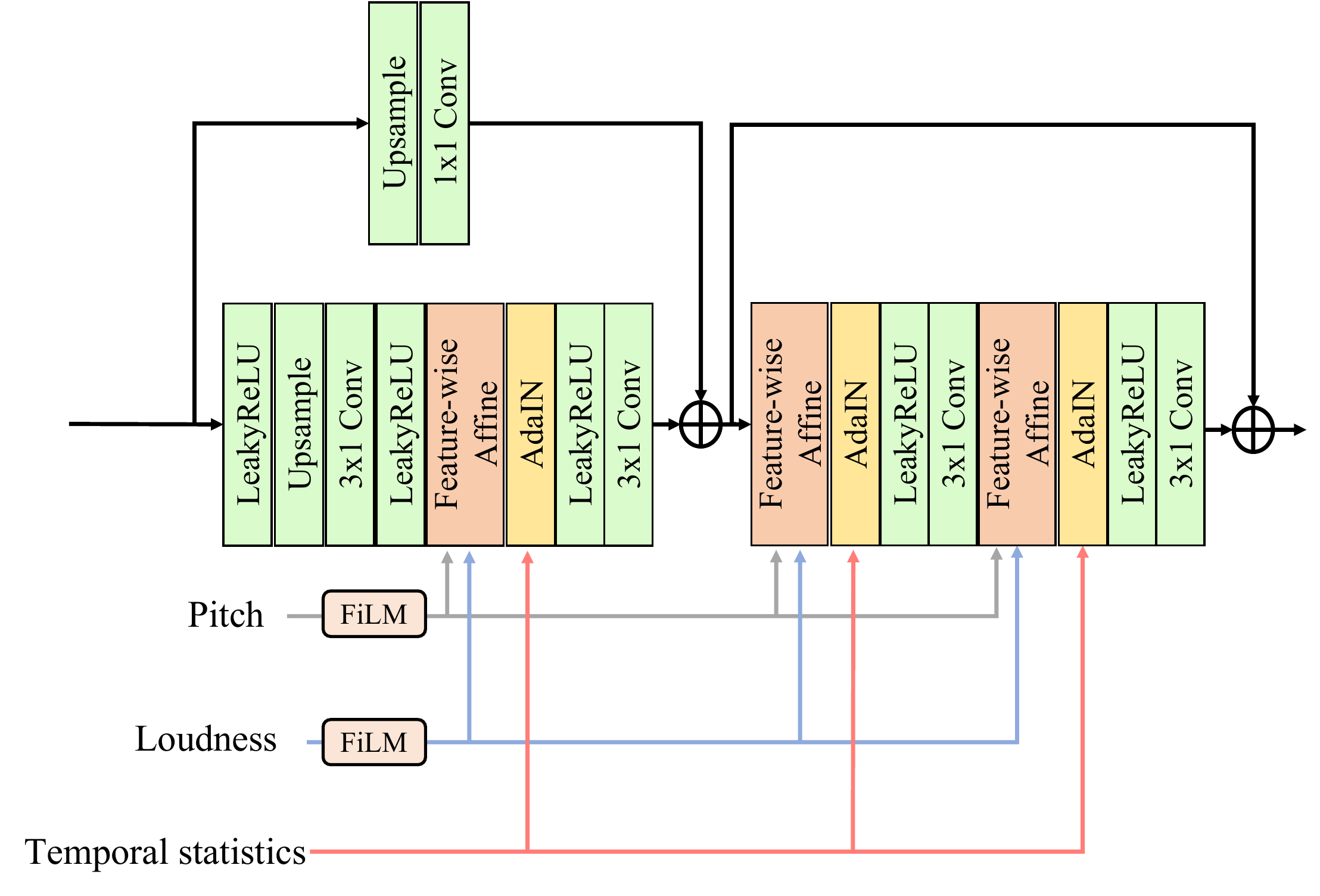}
    \caption{The structure of the up-sampling block, where the ``FiLM" is a feature-wise linear modulation adopted in \cite{liu2021fastsvc,perez2018film}.}
    \label{fig:upsample_block}
\end{figure}

Firstly, the frame-rate F0 values are extracted from the source singing audio by Parselmouth \footnote{https://github.com/YannickJadoul/Parselmouth}, then up-sampled by linear interpolation to match the rate of audio samples, denoted by $f_{1:T}$.
Finally, the harmonic sine-excitation values $F_{it}$ are derived from $f_{1:T}$ as Eq.\ref{eq:sine_excitation}:
\begin{align}
\label{eq:sine_excitation}
    F_{it} = \begin{cases}
     0.1 \sin (\sum_{k=1}^{t} 2\pi \frac{i \times f_{k}}{f_{s}}+\phi_{i}) + n_{it}, & f_{t} > 0 \\
     100 n_{it}, & f_{t} = 0
    \end{cases}
\end{align}
where $i \in \{1, 2, ..., K\}$, $t \in \{1, 2, ..., T\}$ are the indexes of harmonics and time, respectively. $F_{it}$ represents the excitation value of the $i$-th harmonic at time $t$. This work sets $K=8$ to use 8 harmonics. ${f_{s}}$ is the audio sampling rate. $\phi_{i}$ is a random initial phase of the $i$-th harmonic. In unvoiced regions where $f_{t}=0$, we set $F_{it}$ as a Gaussian noise $n_{it} \sim \mathcal{N}(0, 0.003^2)$.

\textbf{Loudness extraction:} This work adopts the A-weighting mechanism of the power spectrum \cite{meyer2005measuring} as the loudness feature.
It assigns greater emphasis to higher frequencies since human ears are less sensitive to low audio frequencies.
The loudness features are computed with the configuration in \cite{hantrakul2019fast}, and the hop-size is set as 64.
It is linearly interpolated to match the rate of audio samples before feeding into the up-sampling stream.

\subsection{The structure of the building blocks}
\label{subsec:wav_gen}
The building blocks are adapted from \cite{liu2021fastsvc}.
We make some modifications to the up-sampling block, while the down-sampling block remains unchanged.
Fig.~\ref{fig:upsample_block} illustrates the structure of the modified up-sampling block, where the feature-wise affine module \cite{chen2020wavegrad} is used to fuse the pitch and the loudness features to the up-sampling block, and the AdaIN \cite{karras2019style} module is fed with the temporal statistics.

In the FiLM module, given one condition feature, i.e. pitch or loudness, it generates a scale vector $\gamma$ and a shift vector $\xi$, which are used to linearly transform the hidden features in the up-sampling block.
Suppose we have the sine-excitation signals $F$, the loudness features $L$ and the hidden features $U$, the feature-wise affine module will transform $U$ into $\tilde{U}$ by Eq.~\ref{eq:feature-wise-affine}:
\begin{align}
    \tilde{U} = (\gamma_F + \gamma_L) \odot U + \xi_F + \xi_L
    \label{eq:feature-wise-affine}
\end{align}
where $\gamma_F$ and $\xi_F$ are the affine parameters associated with $F$, $\gamma_L$ and $\xi_L$ are the ones associated with $L$, and $\odot$ denotes the Hadamard product.

The AdaIN module first normalizes the hidden features by instance normalization, then utilizes the given temporal statistics to linearly transform the hidden features.
This process removes the speaker characteristics of the up-sampling stream, then adapts the hidden features with the new speaker characteristics, i.e. the temporal statistics, generated from the down-sampling stream.
In the conversion stage, the down-sampling stream operates on a target reference audio, then the temporal statistics containing the target speaker characteristics are fused to the up-sampling blocks to perform SVC.
This work adopts all ``IN" modules as the mean normalization and the variance statistics are not computed or used.
Because in our experiments, we observed a quality degradation of the synthesized audio when the mean-variance normalization is applied.
One possible explanation is that the variance statistics could be largely affected by linguistic information, and it may be less helpful when transferring the variance from one audio to another that has mismatched linguistic contents, e.g. at the inference stage.

\subsection{Training loss}
\label{subsec:training-loss}
The system is trained with a combination of three losses: the least-squares GAN \cite{mao2017least} loss, a multi-scale STFT loss \cite{yamamoto2020parallel} and the reconstruction loss on linguistic features.

The discriminator adopts a multi-scale discriminator architecture \cite{kumar2019melgan}.
Given the three sub-discriminators $D_{k}$ ($k \in \{1, 2, 3\}$), the ground-truth audio $x$ and the predicted audio $\hat{x}$, the adversarial loss of the generator can be expressed as:
\begin{align}
    \mathcal{L}_{adv} = \frac{1}{3}\sum_{k=1}^{3}\|1-D_{k}(\hat{x})\|_2
    \label{eq:loss_gen_adv}
\end{align}
The discriminator loss $\mathcal{L}_{D}$ is formulated by Eq.~\ref{eq:loss_dis_adv}:
\begin{align}
    \mathcal{L}_{D} = \frac{1}{3}\sum_{k=1}^{3}(\|1-D_{k}(x)\|_2+\|D_{k}(\hat{x})\|_2)
    \label{eq:loss_dis_adv}
\end{align}
The multi-scale STFT loss is computed as:
\begin{align}
    \mathcal{L}_{stft} = \frac{1}{\lvert M \rvert}\sum_{m \in M}(\frac{\|S_{m}-\hat{S}_{m}\|_2}{\|S_{m}\|_2} + \frac{\|log S_{m} - log \hat{S}_{m}\|_1}{N})
    \label{eq:loss_stft}
\end{align}
where $S_{m}$ and $\hat{S}_{m}$ are the STFT magnitudes computed from $x$ and $\hat{x}$ respectively, with FFT sizes of $m \in M = \{2048, 1024, 512, 256, 128, 64\}$ and with 75\% overlap. $N$ is the total number of elements in $S_{m}$.

The reconstruction loss is the mean squared error between the ground-truth linguistic features $c$ and the predicted ones $\hat{c}$:
\begin{align}
    \mathcal{L}_{mse} = \frac{1}{K}\sum_{i=1}^{K}(c_{i} - \hat{c}_{i})^2
    \label{eq:loss_ppg_rec}
\end{align}
where $K$ is the number of elements in $c$. This constraint guides the down-sampling network to produce speaker-invariant $\hat{c}$.

Finally, the generator loss $\mathcal{L}_G$ is formulated as:
\begin{align}
    \mathcal{L}_{G} = \mathcal{L}_{stft} + \alpha \mathcal{L}_{adv} + \beta \mathcal{L}_{mse}
    \label{eq:loss_gen}
\end{align}
where $\alpha$ and $\beta$ are both experimentally set as 2.5 in this work.

\section{Experimental Setup}
\label{sec:expt-setup}
The proposed model is evaluated in both in-set and out-set conditions.
The in-set evaluation performs SVC among training speakers, while the out-set evaluation alters the voice of singing signals to the voice of out-set speakers which are unseen during training.
The commonly used NUS-48E \cite{duan2013nus} corpus is adopted to train the model and perform the in-set evaluation.
It contains 12 singers with around 15 minutes of data per singer. 
The dataset is randomly divided into train-validation-test sets according to a 90\%-5\%-5\% split. 
For the out-set evaluation, we randomly select 2 male and 2 female singers from the NHSS \cite{sharma2021nhss} corpus, then use the models trained on the NUS-48E corpus to perform SVC to these unseen singers.
All audios are down-sampled to 16kHz in our experiments.

The LUT and the speaker embedding-based SVC systems are adopted as two baselines.
These baselines follow exactly the same configurations as our proposed system, except for the speaker modeling part.
The LUT-based system utilizes a speaker LUT, while the speaker embedding-based system leverages the state-of-the-art speaker recognition model, ECAPA-TDNN \cite{desplanques2020ecapa}, to extract speaker embeddings.
The extracted speaker representations are concatenated with the linguistic features before feeding into the up-sampling block.
The mean squared error in Eq.~\ref{eq:loss_ppg_rec} is dropped in the two baseline models.

In the building blocks, since the linguistic features have a hop-size of 640 samples, 5 up-sampling blocks gradually up-sample the time dimension of the linguistic features by factors of 2, 4, 4, 4, 5 with the number of channels of 288, 192, 96, 48, 24, respectively.
The dilation rates are 1, 3, 9, 27 in all up-sampling blocks.
The ``Conv1D 3$\times$1" in Fig.~\ref{fig:proposed_framework} has 384 channels with dilation rate of 1.
The number of channels of the down-sampling blocks matches those in the corresponding up-sampling blocks. The dilation rates are 1, 2, 4 in all down-sampling blocks.
The LeakyReLU activation function uses a negative slope of 0.2.

This work adopts the ADAM optimizer \cite{kingma2014adam} to train all SVC systems.
Each mini-batch contains 32 audio segments of a one-second length. 
The learning rate is initialized as 0.001 and halved for every 100k training steps.
The discriminator joins the training process after 100k steps.

\section{Experiment Results}
\label{sec:expt-results}

\subsection{Subjective evaluation}
Subjective evaluation is conducted under both in-set and out-set conditions to verify the effectiveness of the proposed system.
The standard 5-scale mean opinion score (MOS) test is adopted to evaluate the naturalness and the voice similarity.
In the naturalness test, each group of stimuli contains recording samples that are randomly shuffled with the ones converted by the compared systems before being presented to raters. 
In the voice similarity test, converted samples are directly compared with the target singers’ reference recordings. 
At least 12 samples are rated for the compared systems in each testing scenario.
We invite 15 Chinese speakers who are also proficient in English to participate in the MOS tests.

Table~\ref{tab:inset-evaluation} illustrates the in-set evaluation results.
The proposed system is observed to perform the best in terms of both naturalness and voice similarity, indicating the effectiveness of the proposed hierarchical speaker modeling with different granularity.
The LUT-based approach outperforms the speaker embedding-based one.
It is possibly because the speaker representations learned by the LUT approach remain constant for all the audios associated with the same training speaker, while the embeddings extracted from ECAPA-TDNN vary across different reference audios.
The LUT approach gives stable voice conversion only for the training speakers, however, it fails to generalize to unseen speakers.
Note that the similarity score variation is larger than that in the naturalness.
This is possibly because the human perception of voice similarity can be largely influenced by pitch values.

The out-set evaluation results are shown in Table~\ref{tab:outset-evaluation}. Since the LUT approach does not support SVC for unseen speakers, we only compare the performance between the proposed system and the speaker embedding-based system.
As shown in Table~\ref{tab:outset-evaluation}, the proposed approach still achieves superior performance to the other system in both naturalness and voice similarity, which verifies the effectiveness of the proposed approach. 


\subsection{Analysis}
\begin{figure}[th]
    \centering
    \includegraphics[width=0.45\textwidth]{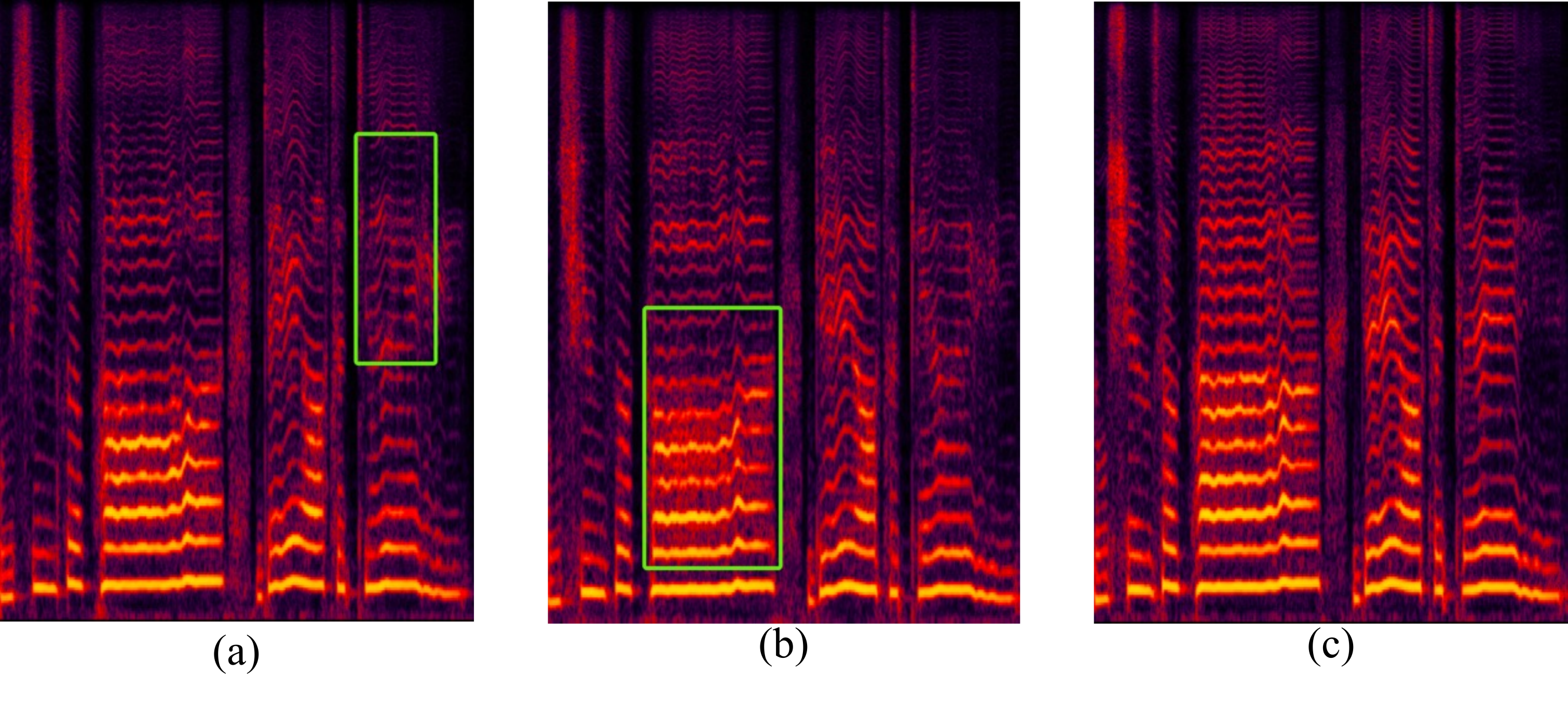}
    \caption{The spectrograms of the converted singing voice by (a) the LUT based system, (b) the speaker embedding based system and (c) the proposed system. (Zoom in for better view.)}
    \label{fig:Spec_compare}
\end{figure}

Fig.~\ref{fig:Spec_compare} shows the spectrograms of the singing voice converted by (a) the LUT-based system, (b) the speaker embedding-based system and (c) the proposed system.
It shows that the spectrogram in (c) has a clearer texture than the ones in (a) and (b).
Specifically, in (a) and (b), the parts in the green box are visibly blurred, which may be caused by the insufficient speaker information within the LUT and the speaker embedding-based systems.
These blurred parts are improved to be clearer in (c).
This demonstrates that the designed hierarchical speaker representation is capable of providing more details to reconstruct the singing voice, which thus boosts the SVC performance.

\begin{table}[t]
\centering
\caption{The in-set evaluation: mean opinion score (MOS) results with 95\% confidence intervals.}
\begin{tabular}{c|c|c}
    \hline
    \hline
                  & Naturalness   & Similarity                 \\
    \hline
    LUT               & 3.43 $\pm$ 0.09 & 3.22 $\pm$ 0.14      \\
    \hline
    ECAPA-TDNN        & 3.36 $\pm$ 0.10 & 3.17 $\pm$ 0.15      \\
    \hline
    Proposed              & \textbf{3.53} $\pm$ 0.09 & \textbf{3.36} $\pm$ 0.12 \\
    \hline
    \hline
    Recordings         & 4.69 $\pm$ 0.07 & 4.39 $\pm$ 0.12   \\
    \hline
    \hline
\end{tabular}
\label{tab:inset-evaluation}
\end{table}

\begin{table}[t]
\centering
\caption{The out-set evaluation: mean opinion score (MOS) results with 95\% confidence intervals.}
\begin{tabular}{c|c|c}
    \hline
    \hline
                      & Naturalness & Similarity \\
    \hline
    LUT                      & --           & --   \\
    \hline
    ECAPA-TDNN      & 3.24 $\pm$ 0.10 & 3.20 $\pm$ 0.16 \\
    \hline
    Proposed            & \textbf{3.35} $\pm$ 0.11 & \textbf{3.29} $\pm$ 0.17 \\
    \hline
    \hline
    Recordings       & 4.77 $\pm$ 0.05 & 4.64 $\pm$ 0.12 \\
    \hline
    \hline
\end{tabular}
\label{tab:outset-evaluation}
\end{table}

\section{Conclusions}
\label{sec:conclusions}
This work proposes a novel SVC framework that extracts a hierarchical speaker representation to capture fine-grained speaker characteristics.
It consists of an up-sampling stream and three down-sampling streams.
The temporal statistics within one down-sampling stream form a hierarchical speaker representation at different granularity, which is fused into the up-sampling block to provide the speaker condition.
Experiment results demonstrate the superior performance of the proposed system over both the LUT and the speaker embedding-based systems.
Moreover, the proposed system supports one-shot SVC with only a few seconds of a target reference audio.

\bibliographystyle{IEEEtran}

\bibliography{mybib}

\end{document}